\documentclass[12pt,a4paper,final]{iopart}
\usepackage{iopams}
\usepackage{float}
\usepackage{graphicx}
\usepackage[breaklinks=true,colorlinks=true,linkcolor=blue,urlcolor=blue,citecolor=blue]{hyperref}
\usepackage{tabularx}

\begin{document}

\title[]{Magnetic structure of  the antiferromagnetic Kondo lattice compounds CeRhAl$_{4}$Si$_{2}$ and CeIrAl$_{4}$Si$_{2}$}
\author{N. J. Ghimire$^{1}$, S. Calder$^{2}$, M. Janoschek$^{1}$,  and E. D. Bauer$^{1}$}
\address{$^{1}$Los Alamos National Laboratory, Los Alamos, New Mexico 87545, USA.
\\\
 $^{2}$ Quantum Condensed Matter Division, Oak Ridge National Laboratory, Oak Ridge, Tennessee 37831, USA.}
\
\ead{nghimire@lanl.gov}

\begin{abstract}

We have investigated the magnetic ground state of the antiferromagnetic Kondo-lattice compounds CeMAl$_{4}$Si$_{2}$ (M = Rh, Ir) using neutron powder diffraction. Although both of these compounds show two magnetic transitions $T_{N1}$ and $T_{N2}$ in the bulk properties measurements, evidence for magnetic long-range order was only found below the lower transition $T_{N2}$. Analysis of the diffraction profiles reveals a commensurate antiferromagnetic structure with a propagation vector $\mathbf{k}$= (0, 0, 1/2). The magnetic moment in the ordered state of CeRhAl$_{4}$Si$_{2}$  and CeIrAl$_{4}$Si$_{2}$ were determined to be 1.14(2) and 1.41(3) $\mu_{B}$/Ce, respectively, and are parallel to the crystallographic $c$-axis in agreement with magnetic susceptibility measurements.
\end{abstract}

\vspace{2pc}


\section[S1]{Introduction}
It is well known that confining the dimensions of a physical system frequently leads to exciting new effects. For example confining electrons to a single dimension results in the breakdown of the Fermi liquid model, and the observation of exotic transport behavior \cite{Schofield1999}. Similarly, there is no long-range magnetic order in two dimensions \cite{Mermin1966}. In $f$-electron materials, where magnetic and electronic degrees of freedom are typically strongly coupled, the effects of dimensionality are currently under debate \cite{Si2001, Abrahams2012, Hackl2011}.  Notably, it has been proposed that dimension can be used to tune the type of observed quantum phase transitions, where either the electronic or magnetic degrees of freedom become quantum critical \cite{Si2001, Coleman10, Si14}.

The entanglement of electronic and magnetic degrees of freedom in $f$-electron compounds is controlled via two magnetic interactions. A large on-site Coulomb repulsion tends to localize the $f$-electron wave function, producing magnetic moments that interact indirectly with each other via the Ruderman-Kittel-Kasuya-Yosida (RKKY) interaction, which leads to a long-range antiferromagnetic order. On the other hand, below a characteristic temperature $T_K$, the Kondo interaction \cite{Kondo64} drives the local demagnetization of the $f$-electron state that is quenched by the spins of the surrounding conduction electrons. This results in delocalization of the $f$-electrons into the conduction band and thus the formation of a paramagnetic heavy Fermi liquid state \cite{Fisk86}.

The most interesting situation occurs when Kondo and RKKY interactions are of comparable size~\cite{Doniach1977}. At this point, the antiferromagnetic ordering temperature is suppressed to zero resulting in a quantum critical point (QCP) where the associated long-wavelength quantum critical magnetic fluctuations are believed to lead to the emergence of novel states of matter~\cite{Loehneysen2007, Gegenwart08}, with the most prominent one being unconventional superconductivity \cite{Mathur1998, Pfleiderer09}. Apart from this prototypical QCP scenario \cite{Hertz, Millis, Moriya}, so-called local quantum criticality has been proposed for the case that the Kondo temperature is suppressed to zero at the QCP, resulting in local critical fluctuations associated with the Fermi surface. In this scenario, the Fermi surface is large on one side of the QCP because it includes the delocalized $f$-electrons, and is small on the other because the $f$-elctrons are localized and do not contribute to the Fermi volume \cite{Si2001}. Recent calculations suggest that the type of QCP may be controlled via the dimensionality of the system \cite{Coleman10,Si14}.

\begin{figure}[h]
\begin{center}
  \includegraphics[scale=1]{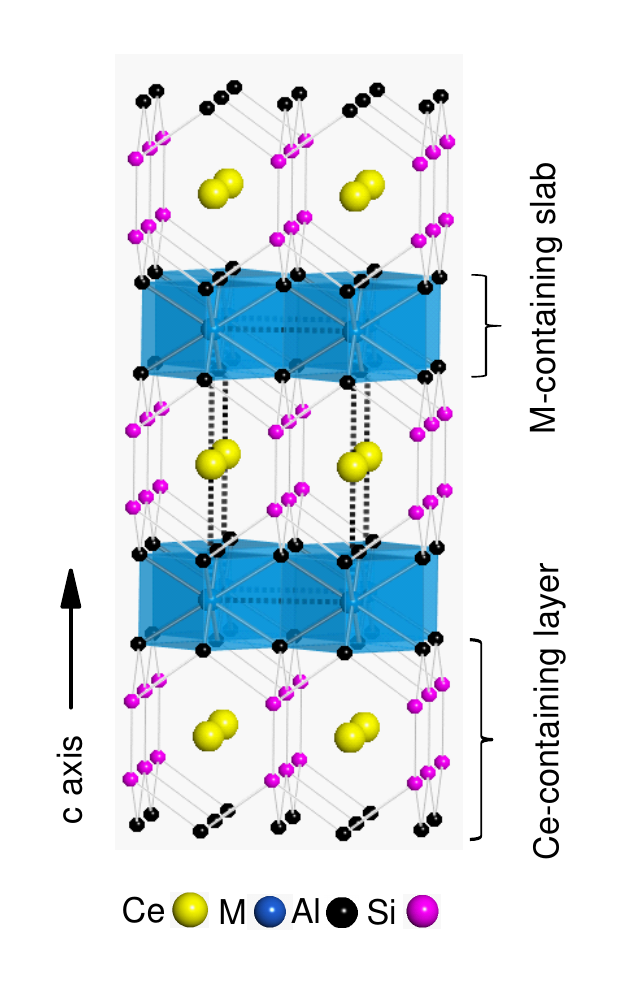}
  \caption{a) Crystal structure of Ce$M$Al$_{4}$Si$_{2}$ (M = Rh and Ir).}\label{fig1}
  \end{center}
\end{figure}

Apart from influencing the type of QCP, lowering the dimensionality is also believed to result in an increase of the superconducting transition temperature \cite{Monthoux2004}. This calls for new classes of $f$-electron materials in which the dimensionality can be systematically tuned. We recently synthesized  Ce$M$Al$_{4}$Si$_{2}$ ($M$ = Rh, Ir, Pt) compounds that are promising in this regard \cite{Ghimire2015}. These materials crystallize in KCu$_{4}$S$_{3}$ structure type with tetragonal space group P4/mmm (\# 123), and exhibit a sequential stacking of a BaAl$_{4}$-type Ce-containing layer separated by a AuAl$_{2}$-type $M$-containing block along the $c$-axis (Fig. \ref{fig1}). Notably, the structure can be described as $n$ = 1 member of Ce$M_{n}$Al$_{2n+2}$Si$_{2}$, where $n$ is an integer. Increasing $n$ enlarges the thickness of the $M$Al$_{2}$ block and hence increases the distance between the Ce planes containing the $4f$ electrons without changing their crystal environment. We note that the large family of $R$(AuAl$_{2}$)$_{n}$Al$_{2}$(Au$_{x}$Si$_{1-x})_{2}$, where $R$ is a rare earth element compounds \cite{Latturner2008} with $n$ = 0 - 3 has already been obtained as a disordered variant (for $x$ = 0.5) of this structural motif ($x$ = 0).

Bulk thermal, transport and magnetic properties reveal two low-temperature antiferromagnetic phases for CeRhAl$_{4}$Si$_{2}$ ($T_{N1}$ = 14 K and $T_{N2}$ = 9 K) and CeIrAl$_{4}$Si$_{2}$ ($T_{N1}$ = 16 K and $T_{N2}$ = 14 K) \cite{Ghimire2015,Maurya2015}. The effective magnetic moment of both CeRhAl$_{4}$Si$_{2}$ and CeIrAl$_{4}$Si$_{2}$ obtained in the paramagnetic region are close to Ce$^{3+}$ Hunds' Rule value (2.54 $\mu_{B}$), indicating localized behavior of the Ce $4f$ moments at high temperature. However, the low temperature behavior point towards screening of these local moments by the Kondo effect \cite{Ghimire2015,Maurya2015}. Heat capacity measurements carried out down to 100 mK determined the Sommerfeld coefficient to be $\gamma$ = 200 mJ/mol-K$^{2}$ in CeRhAl$_{4}$Si$_{2}$ and 50 mJ/mol-K$^{2}$ in CeIrAl$_{4}$Si$_{2}$, indicating  moderately heavy fermion behavior. To make progress on the characterization of these materials, we have carried out neutron powder diffraction (NPD) measurements on CeRhAl$_{4}$Si$_{2}$ and CeIrAl$_{4}$Si$_{2}$ to determine their magnetic structure.

\section[S2]{Experimental Details}

Single crystals of Ce$M$Al$_{4}$Si$_{2}$ ($M$ = Rh, Ir) were grown from Al/Si flux \cite{Ghimire2015}, and ground into fine powders. For CeRhAl$_{4}$Si$_{2}$ 2.5 grams of powder was loaded into a aluminum can with an inner diameter of 6 mm under a He gas atmosphere. In order to reduce the effects of neutron absorption due to Ir, the CeIrAl$_{4}$Si$_{2}$ powder (3.5 g) was loaded into an annular aluminum can, where the annulus was 1 mm thick. The neutron powder diffraction experiments were carried out using the instrument HB2A at the High Flux Isotope Reactor at Oak Ridge National Laboratory. Neutrons of wavelength 2.413 {\AA}, selected by a vertically focusing a Ge(113) monochromator, were used for the experiment. The calculated $1/e$ absorption length for this wavelength is 7.3 mm and 2.5 mm for CeRhAl$_4$Si$_2$ and CeIrAl$_4$Si$_2$, respectively, matching the diameters of the selected aluminum cans. For both compounds, powder diffraction patters were collected for 6 hours for each of the two magnetic phases as well as slightly above the higher temperature magnetic phase. In addition, the temperature variation of the intensity of the strongest magnetic peak was also measured to obtain the temperature dependence of the magnetic order parameter. At each temperature, the peak intensity was measured for three and six minutes for CeRhAl$_{4}$Si$_{2}$ and CeIrAl$_{4}$Si$_{2}$, respectively. Finally, several NPD patterns were collected while cooling from room temperature (100 K) to the base temperature of 3.2 K in the case of CeRhAl$_{4}$Si$_{2}$ (CeIrAl$_{4}$Si$_{2}$) to check for structural distortions. Rietveld refinement \cite{Rietveld} of both the crystal structure and magnetic structure from the NPD data was carried out using the FullProf software \cite{FullProf}. The magnetic structures allowed by symmetry were obtained via a representational analysis using the software SARAh \cite{Wills2000}.

\section[3]{Results and Discussion}

\begin{figure}[h]
\begin{center}
  \includegraphics[scale=1]{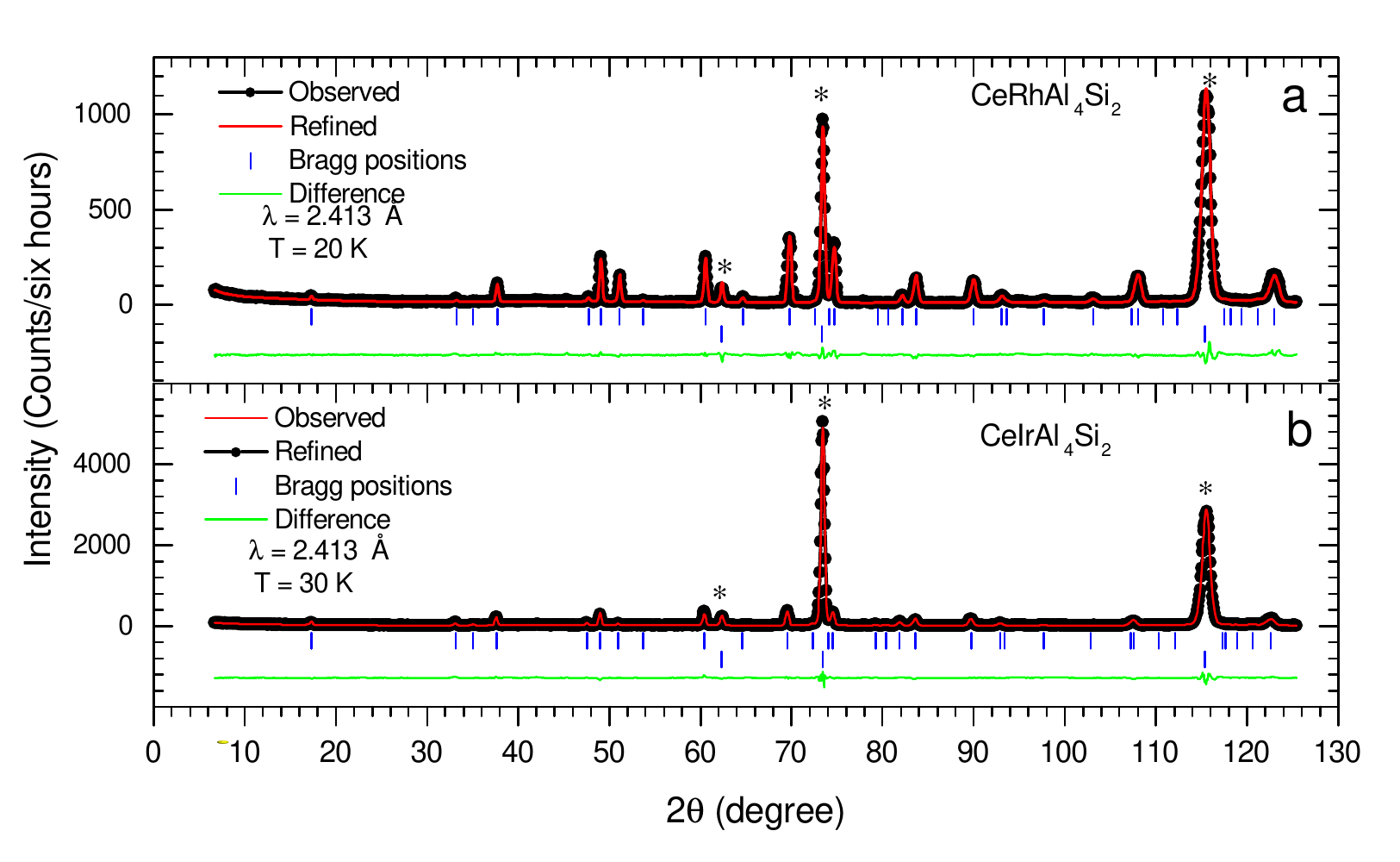}
  \caption{Rietveld refinement of the neutron powder pattern of a) CeRhAl$_{4}$Si$_{2}$ and b) CeIrAl$_{4}$Si$_{2}$ measured above the magnetic transition temperature. The Al peaks from the sample holder are indicated by an asterisk (*), which were modeled by constant profile matching (Le Bail fit).}\label{fig2}
  \end{center}
\end{figure}

Figs. \ref{fig2}(a) and (b) show the Rietveld refinement of the neutron powder diffraction patterns of CeRhAl$_{4}$Si$_{2}$ and CeIrAl$_{4}$Si$_{2}$ collected in the paramagnetic state at $T$ = 20  and 30 K, respectively. The aluminum peaks from the sample holder are marked by asterisks. The results of the refinements are presented in Table \ref{Table1}. In each case, the crystal structure deduced from single crystal x-ray diffraction \cite{Ghimire2015} was used as a starting model. For both samples, no new Bragg peaks appeared during cooling, and within the resolution of the collected data, no indication of a structural distortion was found. Although the samples showed no impurity peaks in the room temperature x-ray patterns, small additional peaks were observed in CeIrAl$_{4}$Si$_{2}$ at all temperatures, suggesting that they can be attributed to a small unknown nonmagnetic impurity phase. We note that the refinement is better in CeRhAl$_{4}$Si$_{2}$ than in CeIrAl$_{4}$Si$_{2}$, most likely because the annular sample can used for CeIrAl$_{4}$Si$_{2}$ may lead to a slight distortion of the peak shapes and intensities.

\begin{figure}[h]
\begin{center}
  \includegraphics[scale=1]{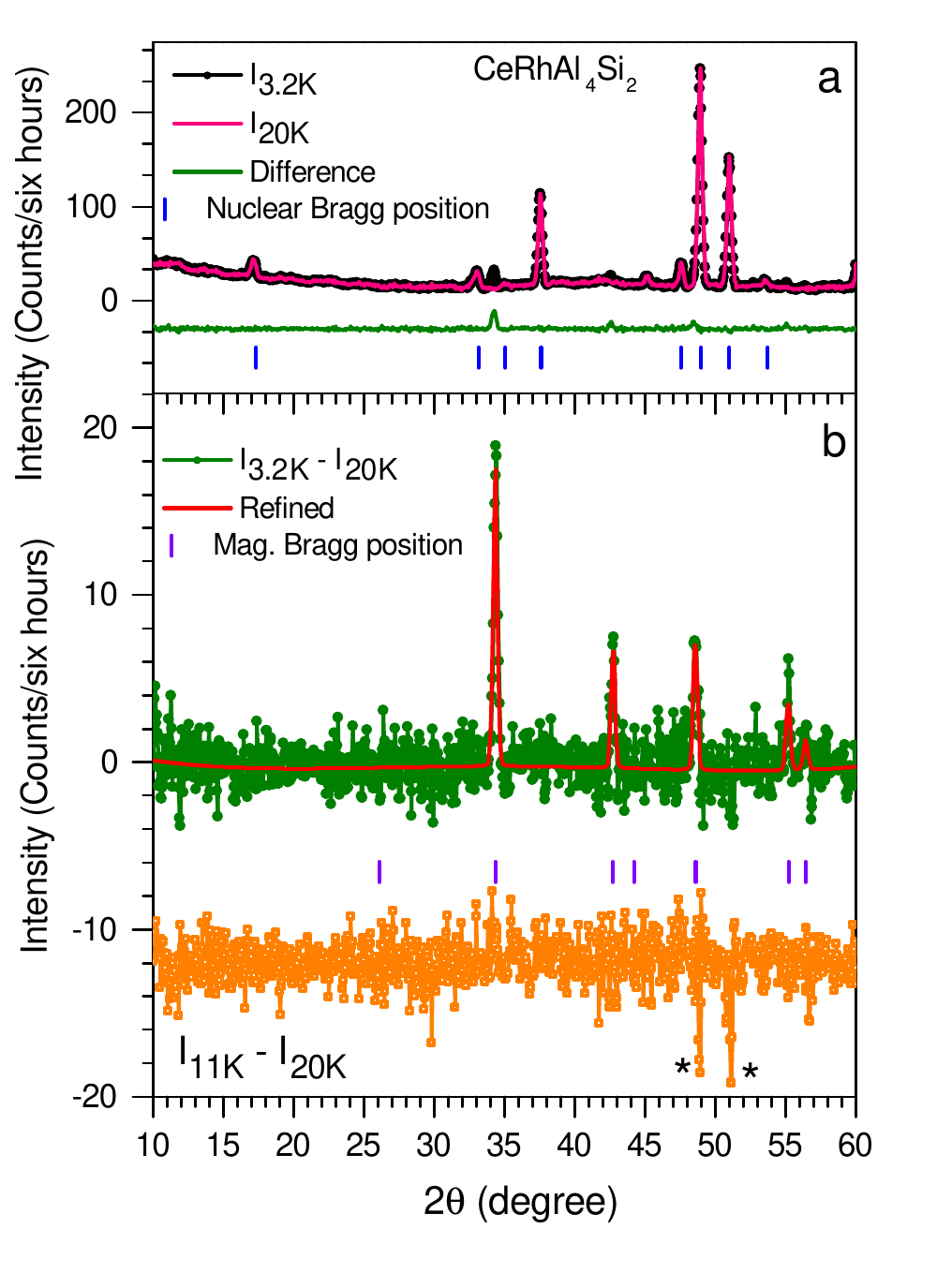}
  \caption{a) Neutron powder diffraction profile of CeRhAl$_{4}$Si$_{2}$ measured at 20 K (pink curve) and at 3.2 K (black curve). The green solid line shows the difference curve obtained from these two profiles.  b) Rietveld refinement of the magnetic peaks at \emph{T} = 3.2 K obtained from the difference pattern $I_{diff}$ = $I_{3.2 K}$ - $I_{20 K}$. The lower (orange) plot is the difference pattern $I_{diff}$ = $I_{11 K}$ - $I_{20 K}$ shifted by 12 below zero. The features marked by asterisks are subtraction artifacts due to the temperature dependence of the Debye-Waller factor. }\label{fig3}
  \end{center}
\end{figure}

\begin{table}
\caption{Crystallographic parameters and agreement factors from Rietveld refinements in space group P4/mmm with Ce atoms at (0, 0, 1/2), M =Rh, Ir at (0, 0, 0), Al atoms at (0, 0, z$_{Al}$) and Si atoms at (1/2, 1/2, z$_{Si}$).}\label{Table1}
\begin{center}
\begin{tabular}{lll}
\hline
	                    &   CeRhAl$_{4}$Si$_{2}$	&	CeIrAl$_{4}$Si$_{2}$	\\
\hline					
T (K)       	            &	20  	                &	30 	                \\
a (\AA)	                &	4.2177(0)	            &	4.2299(2)	            \\
c (\AA)	                &	8.0098(2)	            &	8.0101(3)	             \\
z$_{Al}$	            &	0.1719(5)	            &	0.1690(9)	             \\
z$_{Si}$	            &	0.3546(1)               &	0.361(1)	              \\
R$_{WP}$ (\%)	        &	12.9	                &	11.9	                   \\
$\chi^{2}$	            &	4.32	                &	9.26	                    \\
\hline					
\end{tabular}
\end{center}
\end{table}

\begin{figure}[h]
\begin{center}
  \includegraphics[scale=1]{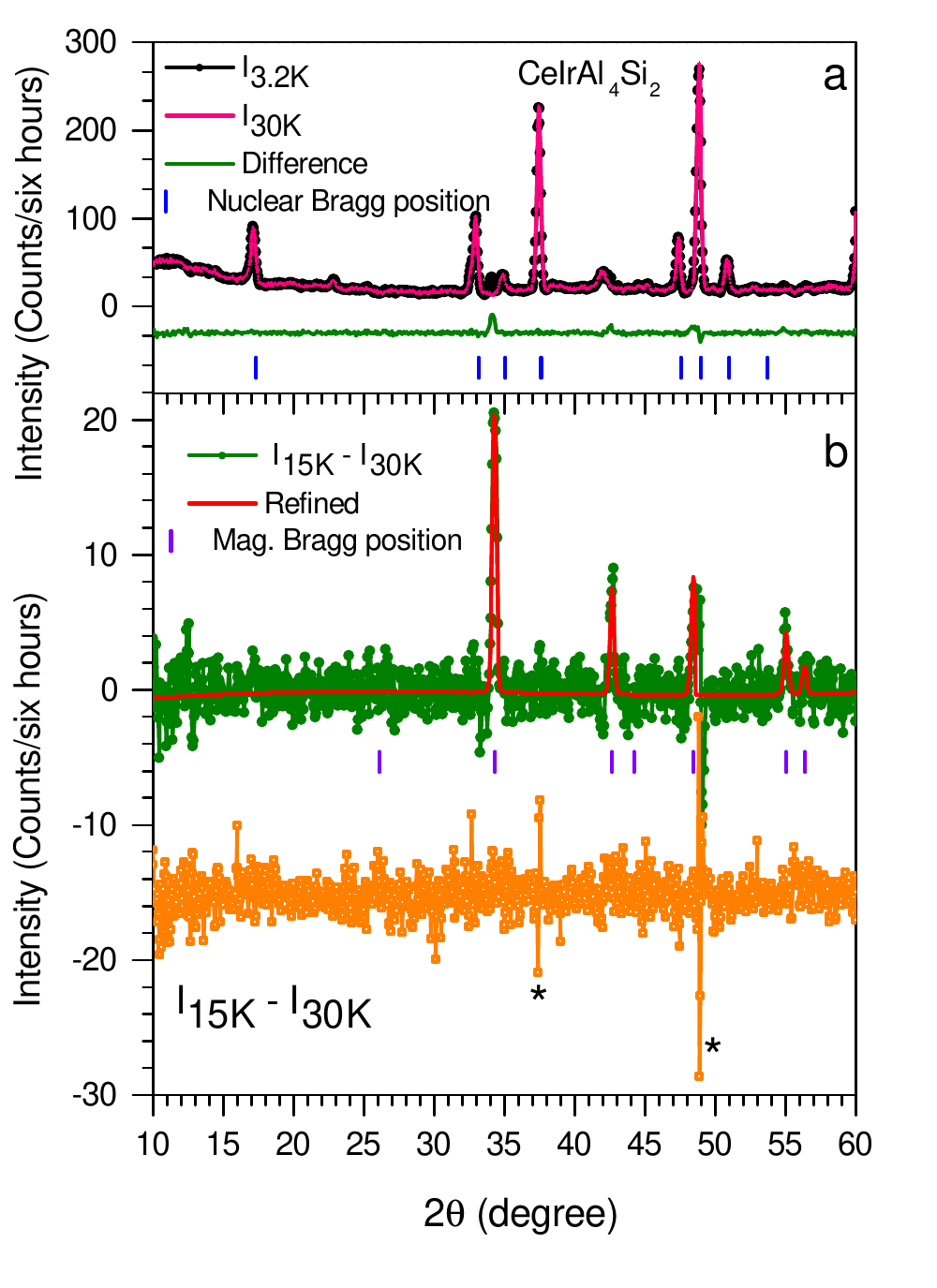}
  \caption{a) Neutron powder diffraction profile of CeIrAl$_{4}$Si$_{2}$ measured at 30 K (pink curve) and at 3.2 K (black curve). Green solid line is the difference of these two profiles. b) Rietveld refinement of the magnetic peaks at \emph{T} = 3.2 K obtained from the difference pattern $I_{diff}$ = $I_{3.2 K}$ - $I_{30 K}$. The lower (orange) plot is the difference pattern $I_{diff}$ = $I_{15 K}$ - $I_{30 K}$ shifted by 15 below zero. The features marked by asterisks are subtraction artifacts due to the temperature dependence of the Debye-Waller factor.}\label{fig4}
  \end{center}
\end{figure}

The low angle portion of the diffraction patterns of CeRhAl$_{4}$Si$_{2}$ collected at $T$ = 20 and 3.2 K are shown in Fig. \ref{fig3}(a). The magnetic contribution to the powder pattern below $T_{N2}$ was obtained by subtracting 20 K dataset from the 3.2 K data, as shown in Fig. \ref{fig3}(b) (green line), and demonstrates the existence of four magnetic Bragg peaks. In contrast, no magnetic Bragg scattering is observed for $T_{N2}<T<T_{N1}$ [orange line in Fig. \ref{fig3}(b)] as revealed by subtracting the 20 K data from a diffraction pattern obtained at $T$ = 11 K. Similar plots of diffraction patterns collected at 30, 15 and 3.2 K for CeIrAl$_{4}$Si$_{2}$ are presented in Fig. \ref{fig4}. Analogous to CeRhAl$_{4}$Si$_{2}$, four magnetic Bragg peaks are observed below $T_{N2}$ for CeIrAl$_{4}$Si$_{2}$, but no sign of long-range magnetic order is detected between $T_{N2}$ and $T_{N1}$.

\begin{table}
\caption{Basis vectors of the space group P4/mmm with a magnetic propagation vector $\mathbf{k}$ = (0, 0, 1/2). The Ce atoms occupy the 1b Wyckoff site in the unit cell, which has only one special position (0, 0, 1/2). $m_{\parallel a}$, $m_{\parallel b}$ and $m_{\parallel c}$ denote the magnetic moment of the Ce atom parallel to $a$, $b$ and $c$-axis, respectively.}\label{Table2}
\begin{center}
\begin{tabular}{ccccc}
\hline
IR	 &   Atom	&	$m_{\parallel a}$  & $m_{\parallel b}$	&  $m_{\parallel c}$     \\
\hline					
$\Gamma_{2}$       	         &    1     &	     0           &	    0           &	    1              \\
$\Gamma_{10}$       	         &    1     &	     0           &	    1           &	    0              \\
                	         &    2     &	     1           &	    0           &	    0              \\
\hline					
\end{tabular}
\end{center}
\end{table}

\begin{figure}[h]
\begin{center}
  \includegraphics[scale=.5]{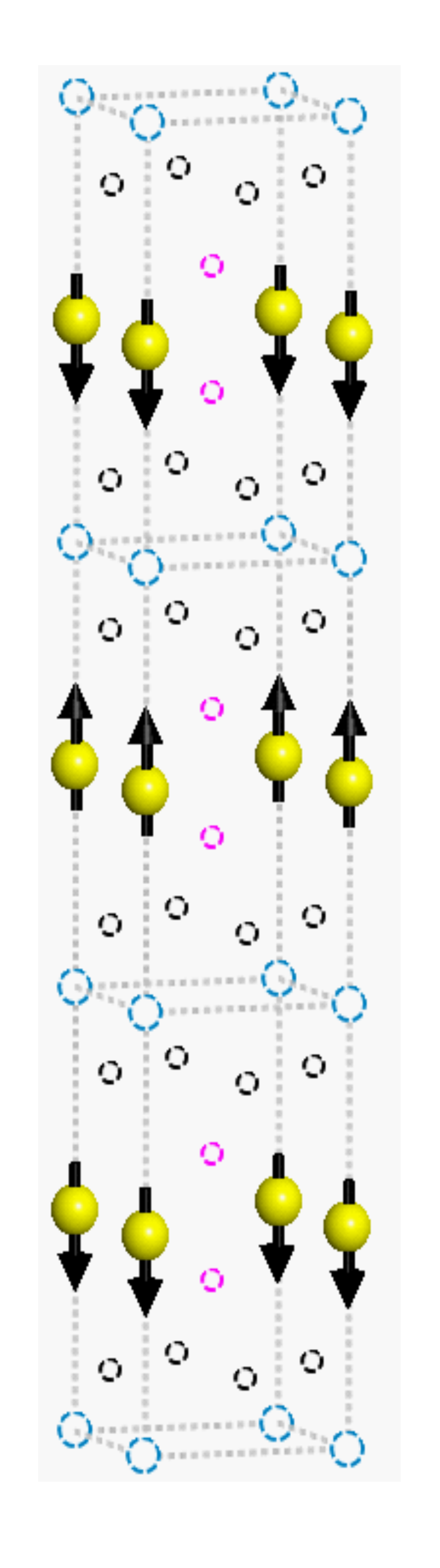}
  \caption{Magnetic structure of CeMAl$_{4}$Si$_{2}$ at \emph{T} = 3.2 K determined from the neutron powder diffraction experiment. The solid (yellow) balls are the Ce atoms and the arrow represents the direction of the magnetic moment.}\label{fig5}
  \end{center}
\end{figure}

The four magnetic Bragg peaks observed in CeMAl$_{4}$Si$_{2}$ (M = Rh, Ir) for temperatures $T<T_{N2}$ may be successfully indexed with a magnetic propagation vector $\mathbf{k}$ = (0, 0, 1/2), indicating a doubling of the magnetic unit cell along the crystallographic $c$-axis compared to the structural unit cell. Using representational analysis based on the identified propagation vector and the space group P4/mmm yields two magnetic irreducible representations (IR) for the Ce site that are summarized in Table \ref{Table2}. Here, the  $\Gamma{_2}$ and $\Gamma_{10}$ irreducible representations describe solutions with magnetic moments parallel and perpendicular to the $c$-axis, respectively. A Rietveld refinement was carried out on the magnetic contribution to the diffraction patterns recorded at $T$ = 3.2 K for CeRhAl$_{4}$Si$_{2}$ and CeIrAl$_{4}$Si$_{2}$, as shown in Figs. \ref{fig3}(b) and \ref{fig4}(b) (red lines), respectively. In agreement with magnetic susceptibility measurements that indicate easy-axis anisotropy along the $c$-axis \cite{Ghimire2015,Maurya2015}, the data is best described by the $\Gamma{_2}$  IR, in which ferromagnetic sheets of Ce magnetic moments are stacked antiferromagnetically along $c$. The corresponding magnetic structure is illustrated in Fig. \ref{fig5}.

\begin{figure}[h]
\begin{center}
  \includegraphics[scale=.8]{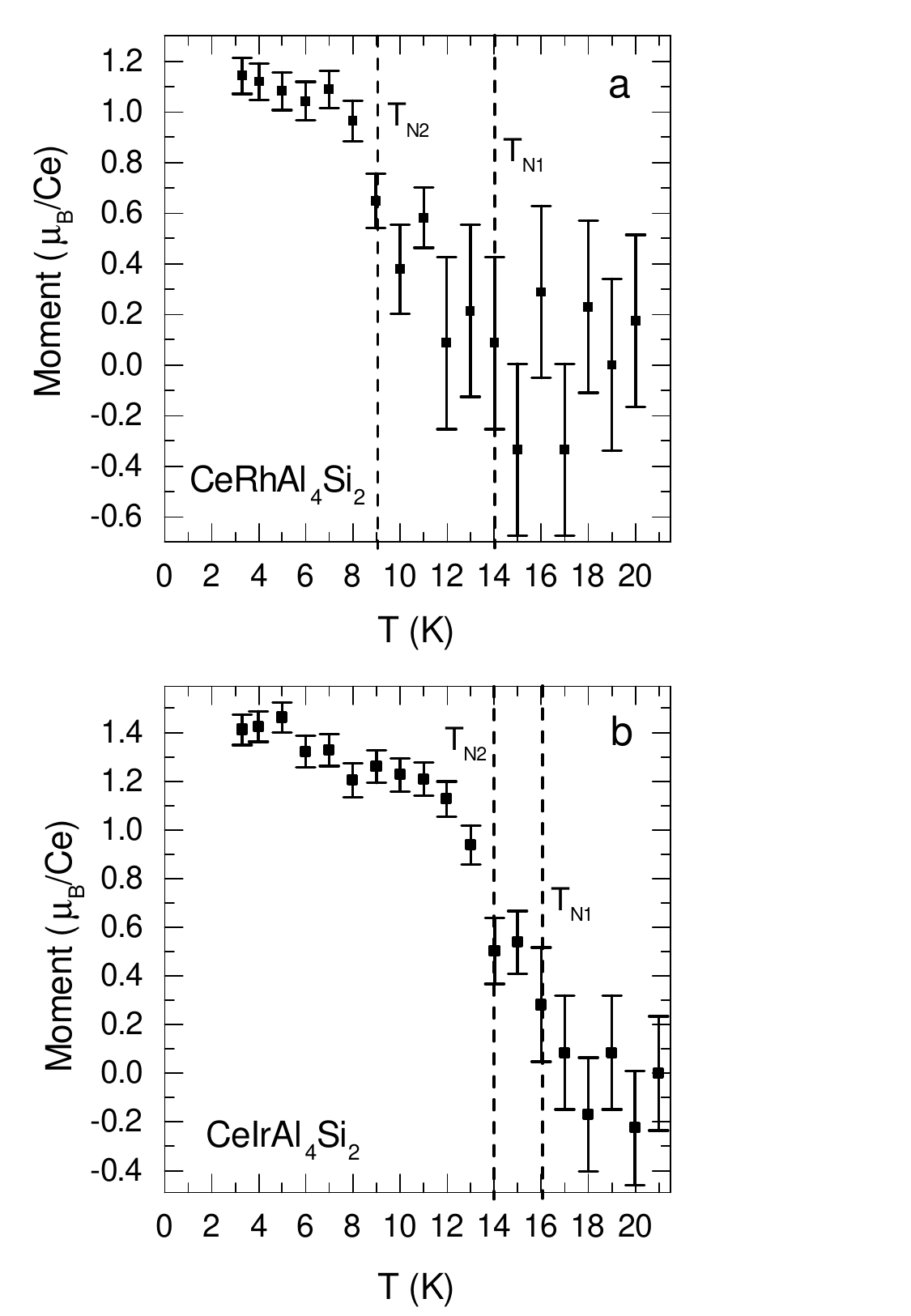}
  \caption{Temperature variation of the magnetic moments of a) CeRhAl$_{4}$Si$_{2}$ and b) CeIrAl$_{4}$Si$_{2}$. Magnetic moments at 3.2 K were obtained from Rietvield refinement of the magnetic Bragg peaks as discussed in the text. The magnetic moments at all other temperatures were obtained by scaling the moment determined at 3.2 K with the square root of the intensity of the strongest peak measured at the respective temperatures.}\label{fig6}
  \end{center}
\end{figure}

The Rietveld fit yields ordered moments of $\mu_{0}$ = 1.14(2) and 1.41(3) $\mu_{B}$/Ce for  CeRhAl$_{4}$Si$_{2}$ and  CeIrAl$_{4}$Si$_{2}$, respectively. The temperature dependence of the staggered magnetization for both compounds was obtained on the strongest magnetic peak and is shown in Fig. \ref{fig6}. The onset of long-range antiferromagnetic order occurs below $T_{N2}$, as signaled by the non-zero magnetic moment.

In summary, the ground state magnetic structure in both CeRhAl$_{4}$Si$_{2}$ and  CeIrAl$_{4}$Si$_{2}$ at 3.2 K was determined to be a commensurate collinear antiferromagnetic structure with propagation vector $\mathbf{k}$ = (0, 0, 1/2) via neutron powder diffraction. Although two magnetic transitions were clearly found in bulk properties measurements, no magnetic Bragg peaks were observed  between $T_{N1}$ and $T_{N2}$.  Possible reasons are that the moments in the region $T_{N2}<T<T_{N1}$ are too small to be observed within neutron powder diffraction, or that the magnetic structure of the high temperature magnetic phase is of an incommensurate nature. We note that both CeRhAl$_{4}$Si$_{2}$ and  CeIrAl$_{4}$Si$_{2}$ contain elements with relatively high neutron absorption, which makes the observation of small magnetic moments difficult. Methods that are typically more sensitive to small ordered magnetic moments such as single crystal neutron diffraction or nuclear magnetic resonance measurements will be required to determine the nature of the magnetic order in between $T_{N1}$ and $T_{N2}$.

The size of the magnetic moments of $\mu$= 1.14(2) and 1.41(3) $\mu_{B}$/Ce for CeRhAl$_{4}$Si$_{2}$ and CeIrAl$_{4}$Si$_{2}$, respectively,  obtained here are consistent with the saturated values of magnetic moments determined via magnetization measurements in the field polarized state (0.95 and 1.14 $\mu_{B}$/Ce for  CeRhAl$_{4}$Si$_{2}$ and  CeIrAl$_{4}$Si$_{2}$, respectively) \cite{Maurya2015}. Moreover, the magnitude of the ordered magnetic moments are much smaller than
than the moment size of a free Ce$^{3+}$ ion given by $g_{J}J$ = 2.14 $\mu_{B}$. Typically, a reduction of the ordered moment in rare earth systems may be due to crystal field effects and/or to Kondo screening. Maurya \emph{et. al.} \cite{Maurya2015} suggested that crystal field effects alone cannot describe the reduced size of the magnetic moment and attributed the reduction of the moment to the Kondo screening of the $4f$ moments. This is supported by heat capacity measurements \cite{Ghimire2015,Maurya2015}, which indicate that the Sommerfeld coefficient is large in CeRhAl$_{4}$Si$_{2}$ ($\gamma =200$ mJ/molk-K$^2$), and moderately enhanced in CeIrAl$_{4}$Si$_{2}$ ($\gamma =50$ mJ/molk-K$^2$), implying that Kondo screening is indeed an important effect in these compounds. We note that ultimately a determination of the crystal electric field parameters by neutron or x-ray scattering will be useful to elucidate the relative roles of Kondo and crystal fields effects on the size of the ordered moment.

Assuming that the Kondo effect is indeed the main contributor in reducing the size of the magnetic moment, both CeRhAl$_{4}$Si$_{2}$ and CeIrAl$_{4}$Si$_{2}$ appear to be promising candidates to investigate the interplay of electronic and magnetic degrees of freedom near a QCP while tuning the dimensionality of the magnetic interactions. Because of its lower magnetic ordering temperature and larger Sommerfeld coefficient (i.e. larger Kondo interaction), CeRhAl$_{4}$Si$_{2}$ is likely to be nearer to a putative antiferromagnetic QCP. Therefore measurements of its magnetic properties as function of chemical substitution or external pressure are highly desirable in order to test whether these novel compounds can be tuned towards a QCP.

\section{Acknowledgements}
Work at Los Alamos National Laboratory was performed
under the auspices of the US Department of Energy, Office
of Basic Energy Sciences, Division of Materials Sciences and
Engineering. Research conducted at ORNL's High Flux Isotope Reactor was sponsored by the Scientific User Facilities Division, Office of Basic Energy Sciences, US DOE.

\section*{References}
\bibliographystyle{ieeetr}

\begin{thebibliography}{10}

\bibitem{Schofield1999}
Schofield A J 1999 {\em Contemporary Physics} {\bf 40} 95.

\bibitem{Mermin1966}
Mermin N D and Wagner H 1966 {\em Phys. Rev. Lett.} {\bf 17} 1133.

\bibitem{Si2001}
Si Q, Rabello S, Ingersent K and Lleweilun Smith J 2001 {\em Nature} {\bf 413} 804.

\bibitem{Abrahams2012}
Abrahams E and Wölfle P 2012 {\em PNAS} {\bf 109} 3238.

\bibitem{Hackl2011}
Hackl A H and Vojta M 2011 {\em Phys. Rev. Lett.} {\bf 106} 137002.

\bibitem{Coleman10}
  Coleman P and Nevidomskyy A H 2010
  {\em J. Low Temp. Phys.} {\bf 161} 182 .

\bibitem{Si14}
 Si Q, Pixley J H, Nica E, Yamamoto S J, Goswami P, Yu R and Kirchner S 2014
  {\em J. Phys. Soc. Jpn.} {\bf 83} 061005.

\bibitem{Kondo64}
  Kondo J 1964
  {\em Prog. Theor. Phys.} {\bf 32} 37.

\bibitem{Fisk86}
  Fisk Z, Ott H R, Rice T M and Smith J L 1986
  {\em Nature} {\bf 320}, 124.

\bibitem{Doniach1977}
Doniach S 1977 {\em Physica} {\bf 91B} 231.

\bibitem{Loehneysen2007}
L\"{o}hneysen H, Rosch A, Vojta, M and W\"{o}lfle P 2007 {\em Rev. Mod. Phys.} {\bf79} 1015.

\bibitem{Gegenwart08}
  Gegenwart P, Si Q and Steglich F 2008
  {\em Nat. Phys.} {\bf 4} 892.

\bibitem{Mathur1998}
Mathur N D, Grosche F M, Julian S R, Walker I R, Freye D M, Haselwimmer R K W and Lonzarich G G 1998 {\em Nature} {\bf 394} 39.

\bibitem{Pfleiderer09}
  Pfleiderer C, 2009
  {\em Rev. Mod. Phys.} {\bf 81} 1551.

\bibitem{Hertz}
Hertz J 1976 {\em Phys Rev B} {\bf 14} 1165.

\bibitem{Millis}
Millis A 1993 {\em Phys Rev B} {\bf 48} 7183.

\bibitem{Moriya}
Moriya T 1985 Spin Fluctuations in Itinerant Electron Systems (Springer, Berlin).

\bibitem{Monthoux2004}
Monthoux P and Lanzarich G G 2004 {\em Phys. Rev. B} {\bf 69} 064517.

\bibitem{Ghimire2015}
Ghimire N J, Ronning F, Williams D J, Scott B L, Lou Y, Thompson J D and
  Bauer E D 2015 {\em J. Phys.: Condens. Matt.} {\bf 27} 025601.

\bibitem{Latturner2008}
Latturner S E and Kanatzidis M G 2008 {\em Inorg. Chem.} {\bf 47} 2089

\bibitem{Maurya2015}
Maurya A, Kulkarni R, Thamizhavel A, and
  Dhar S K 2015 arXiv:1501.00250.

\bibitem{Rietveld}
McCusker L B, Von Dreele R B, Cox D E, Lou\"{e}r D and Scardi P 1999 {\em J. Appl. Cryst.} {\bf 32} 36.

\bibitem{FullProf}
Rodriguez-Carvajal J 1993 {\em Physica B} {\bf 192} 55.

\bibitem{Wills2000}
Wills A S 2000 {\em Physica B} {\bf 276} 680.

\bibitem{Bertaut}
Bertaut E F 1968 {\em Acta Cryst.} {\bf A24} 217.


\end{thebibliography}

\end{document}